\documentclass[aps,pra,twocolumn,showpacs]{revtex4}
\usepackage{amsmath}
\usepackage{graphicx}
\usepackage{amsfonts}
\usepackage{amssymb}
\begin{document}
\title{Locally accessible information from multipartite ensembles}
\author{Wei Song}
\affiliation{Hefei National Laboratory for Physical Sciences at
Microscale and Department of Modern Physics, University of Science
and Technology of China, Hefei, Anhui 230026, China}

\pacs{03.67.Mn, 03.65.Ud, 03.67.Hk}

\begin{abstract}
We present a universal Holevo-like upper bound on the locally
accessible information for arbitrary multipartite ensembles. This
bound allows us to analyze the indistinguishability of a set of
orthogonal states under LOCC. We also derive the upper bound for the
capacity of distributed dense coding with multipartite senders and
multipartite receivers.

\end{abstract}
\maketitle

It is well known that any set of orthogonal states can be
discriminated if there are no restrictions to measurements that one
can perform. However, discrimination with certainty is not
guaranteed for multipartite orthogonal states, if only local
operations and classical communication (LOCC) are allowed
\cite{Ghosh:2001,Walgate:2000,Chen:2003,Fan:2004,Watrous:2005,Ye:2007}.
For example, more than two orthogonal Bell states with a single copy
cannot be distinguished by LOCC\cite{Ghosh:2001}. In Ref.
\cite{Bennett:1999} Bennett \emph{et al} constructed a set of
orthogonal bipartite pure product states, that cannot be
distinguished with certainty by LOCC. Another counterintuitive
result was obtained in Ref. \cite{Horodecki:2003}: there are
ensembles of locally distinguishable orthogonal states, for which
one can destroy local distinguishability by reducing the average
entanglement of the ensemble states. To understand these interesting
results deeply, it is important to investigate the connection
between classical and quantum information and extraction of
classical information about the ensemble by local operations and
classical communication.

An important step is made in Ref.\cite{Badziag:2003}, Badzig
\emph{et al}. found a universal Holevo-like upper bound on the
locally accessible information. They show that for a bipartite
ensemble $\left\{ {p_x ,\rho _x^{AB} } \right\}$, the locally
accessible information is bounded by

\begin{equation}
\label{eq1} I^{LOCC} \le S\left( {\rho ^A} \right) + S\left( {\rho
^B} \right) - \mathop {\max }\limits_{Z = A,B} \sum\limits_x {p_x
S\left( {\rho _x^Z } \right)} ,
\end{equation}

\noindent where $\rho ^A$ and $\rho ^B$ are the reductions of
$\rho ^{AB} = \sum\nolimits_x {p_x \rho _x^{AB} } $, and $\rho
_x^Z $ is a reduction of $\rho _x^{AB} $.

In this paper, we will prove a multipartite generalization of this
bound. First we consider an arbitrary tripartite ensemble $R =
\left\{ {p_x ,\rho _x^{ABC} } \right\}$ to give an example. The
central tool we will require is the following
result\cite{Badziag:2003}, which is a generalization of the Holevo
bound on mutual information.

\emph{Lemma 1}. If a measurement on ensemble $Q = \left\{ {p_x ,\rho
_x } \right\}$ produces result $y$ and leaves a postmeasurement
ensemble $Q^y = \left\{ {p_{x\vert y} ,\rho _{x\vert y} } \right\}$
with probability $p_y $, then the mutual information $I$ (between
the indentity of state in the ensemble and measurement outcome)
extracted from the measurement is bounded by

\begin{equation}
\label{eq2} I \le \chi _Q - \bar {\chi }_{Q^y} ,
\end{equation}

\noindent where $\bar {\chi }_{Q^y} $ is the average Holevo bound
for the possible postmeasurement ensemble, i.e., $\sum\nolimits_y
{p_y \chi _{Q^y} } $. Suppose that Alice, Bob and Charlie are far
apart and the allowed measurements strategies are limited to
LOCC-based measurements. Without loss of generality, let Alice make
the first measurement, and suppose that she obtains an outcome $a$
with probability $p_a $. Suppose that the postmeasurement ensemble
is $R_a = \left\{ {p_{x\vert a} ,\rho _{x\vert a}^{ABC} } \right\}$.
Lemma 1 bounds the mutual information obtained from Alice as
follows: $I_1^A \le \chi _{R^A} - \bar {\chi }_{R_a^A } ,$ where
$\chi _{R^A} $ is the Holevo quantity of the $A$ part of the
ensemble $R$ and $\chi _{R_a^A } $ is the Holevo quantity of the $A$
part of the ensemble $R_a $. After Bob has learned the Alice's
result was $a$, his ensemble is denoted by $R_a^B = \left\{
{p_{x\vert a} ,\rho _{x\vert a}^B } \right\}$, with $\rho _x^B =
tr_{AC} \left( {\rho _x^{ABC} } \right)$, Suppose Bob performs the
second measurement and obtains outcome $b$ with probability $p_b $,
then the postmeasurement ensemble is $R_{ab} = \left\{ {p_{x\vert
ab} ,\rho _{x\vert ab}^{ABC} } \right\}$. Using Lemma 1, the mutual
information obtained from Bob's measurement has the bound: $I_2^B
\le \bar {\chi }_{R_{^a}^B } - \bar {\chi }_{R_{ab}^B } ,$ where
$\bar {\chi }_{R_{^a}^B } = \sum\nolimits_a {p_a \left[ {S\left(
{\sum\nolimits_x {p_{x\vert a} \rho _{x\vert a}^B } } \right) -
\sum\nolimits_x {p_{x\vert a} S\left( {\rho _{x\vert a}^B } \right)}
} \right]} $, and $\bar {\chi }_{R_{^{ab}}^B } = \sum\nolimits_{ab}
{p_{ab} \left[ {S\left( {\sum\nolimits_x {p_{x\vert ab} \rho
_{x\vert ab}^B } } \right) - \sum\nolimits_x {p_{x\vert ab} S\left(
{\rho _{x\vert ab}^B } \right)} } \right]} $. Similarly, the
information extracted from Charlie's measurement is bounded as
follows: $I_3^C \le \bar {\chi }_{R_{^{ab}}^C } - \bar {\chi
}_{R_{abc}^C } ,$ where we have assumed that Charlie obtains an
outcome $c$ with probability $p_c $. This procedure goes for an
arbitrary number of steps, thus the total information gathered from
all steps is $I^{LOCC} = I_1^A + I_2^B + I_3^C + \cdots $, where the
subscript $n$ denotes the information is extracted from the $n$th
measurement. To proceed with our derivations, we need the following
facts: \newline \textbf{(i)}Concavity of the von Neumann
entropy.\newline \textbf{(ii)}A measurement on one subsystem does
not change the density matrix at a distant subsystem.\newline
\textbf{(iii)}A measurement on one subsystem cannot reveal more
information about a distant subsystem than about the subsystem
itself. For example, after the first measurement by Alice, we have
$\sum\nolimits_x {p_x S\left( {\rho _x^A } \right)} -
\sum\nolimits_a {p_a \sum\nolimits_x {p_{x\vert a} S\left( {\rho
_{x\vert a}^A } \right)} }  \ge \sum\nolimits_x {p_x S\left( {\rho
_x^B } \right)} - \sum\nolimits_a {p_a \sum\nolimits_x {p_{x\vert a}
S\left( {\rho _{x\vert a}^B } \right)} } $.

Suppose that the last measurement is performed by Alice, then
after $n$ steps of measurements, we obtain the following
inequality

\begin{equation}
\label{eq3}
\begin{array}{l}
 I^{LOCC} \le S\left( {\rho ^A} \right) + S\left( {\rho ^B} \right) +
S\left( {\rho ^C} \right) - \sum\nolimits_x {p_x S\left( {\rho
_x^C }
\right)} \\
 - \sum\nolimits_{a,b,\ldots ,n} {p_{a,b,\ldots ,n} S\left( {\sum\nolimits_x
{p_{x\vert a,b,\ldots ,n} \rho _{x\vert a,b,\ldots ,n}^A } } \right)} \\
 \end{array},
\end{equation}

\noindent where $\left\{ {p_{x\vert a,b,\ldots ,n} ,\rho _{_{x\vert
a,b,\ldots ,n} }^{ABC} } \right\}$ is the postmeasurement ensemble
obtained after the measurement in the $n$th step and $p_{a,b,\ldots
,n} $ is the probability of the sequence of measurement in steps
$1,2,\ldots ,n$. If the last measurement is performed by Bob. We
have

\begin{equation}
\label{eq4}
\begin{array}{l}
 I^{LOCC} \le S\left( {\rho ^A} \right) + S\left( {\rho ^B} \right) +
S\left( {\rho ^C} \right) - \sum\nolimits_x {p_x S\left( {\rho
_x^A }
\right)} \\
 - \sum\nolimits_{a,b,\ldots ,\left( {n + 1} \right)} {p_{a,b,\ldots ,\left(
{n + 1} \right)} S\left( {\sum\nolimits_x {p_{x\vert a,b,\ldots
,\left( {n + 1} \right)} \rho _{x\vert a,b,\ldots ,\left( {n + 1}
\right)}^B } } \right)}
\\
 \end{array}.
\end{equation}

When the last measurement is performed by Charlie, the inequality
takes the form

\begin{equation}
\label{eq5}
\begin{array}{l}
 I^{LOCC} \le S\left( {\rho ^A} \right) + S\left( {\rho ^B} \right) +
S\left( {\rho ^C} \right) - \sum\nolimits_x {p_x S\left( {\rho
_x^B }
\right)} \\
 - \sum\nolimits_{a,b,\ldots ,\left( {n + 2} \right)} {p_{a,b,\ldots ,\left(
{n + 2} \right)} S\left( {\sum\nolimits_x {p_{x\vert a,b,\ldots
,\left( {n + 2} \right)} \rho _{x\vert a,b,\ldots ,\left( {n + 2}
\right)}^C } } \right)}
\\
 \end{array}.
\end{equation}

The last terms in Eqs.(\ref{eq3})-(\ref{eq5}) are all negative
values. Neglecting these terms, we have

\begin{equation}
\label{eq6} I^{LOCC} \le S\left( {\rho ^A} \right) + S\left( {\rho
^B} \right) + S\left( {\rho ^C} \right) - \mathop {\max
}\limits_{Z = A,B,C} \sum\nolimits_x {p_x S\left( {\rho _x^Z }
\right)} .
\end{equation}

For a multipartite ensembles more than three components we can prove
the following Lemma by the same way as proving the above results.

\emph{Lemma 2}. For an arbitrary multipartite ensemble $\left\{ {p_x
,\rho _{_x }^{B_1 B_2 \cdots B_N } } \right\}$, the maximal locally
accessible mutual information satisfies the inequality:

\begin{equation}
\label{eq7}
\begin{array}{l}
 I^{LOCC} \le S\left( {\rho ^{B_1 }} \right) + S\left( {\rho ^{B_2 }}
\right) +  \cdots + S\left( {\rho ^{B_N }} \right) \\
 - \mathop {\max }\limits_{Z = B_1 ,B_2 ,\ldots ,B_N } \sum\nolimits_x {p_x
S\left( {\rho _x^Z } \right)} \\
 \end{array},
\end{equation}

\noindent where $\rho ^{B_n }$ is the reduction of $\rho ^{B_1 ,B_2
,\ldots ,B_N } = \sum\nolimits_x {p_x \rho _x^{^{B_1 ,B_2 ,\ldots
,B_N }} } $and $\rho _x^{^{Z}} $ is a reduction of $\rho _x^{^{B_1
,B_2 ,\ldots ,B_N }} $.

While the ensemble states $\rho _{_x }^{B_1 B_2,\ldots ,B_N } $ are
all pure states, it is possible to write Eq.(\ref{eq7}) in a form of
the average multipartite q-squashed entanglement. Notice that for
the N-partite pure state $\left| \Gamma \right\rangle _{A_1 ,\ldots
,A_N } $, we have \cite{Yang:2007}$E_{sq}^q \left( {\left| \Gamma
\right\rangle _{A_1 ,\ldots ,A_N } } \right) = S\left( {\rho _{A_1 }
} \right) + \cdots + S\left( {\rho _{A_N } } \right)$, where $\rho
_{A_k } = Tr_{A_1 ,\ldots ,A_{k - 1} ,A_{k + 1} ,\ldots ,A_N }
\left( {\left| \Gamma \right\rangle \left\langle \Gamma \right|}
\right)$, then Eq.(\ref{eq7}) can be rewritten as $I^{LOCC} \le
S\left( {\rho ^{B_1 }} \right) + S\left( {\rho ^{B_2 }} \right) +
\cdots + S\left( {\rho ^{B_N }} \right) - \sum\nolimits_x {p_x
\frac{E_{sq}^q \left( {\left| \psi \right\rangle _{_x }^{B_1 B_2
,\ldots ,B_N } } \right)}{N}} $, where $\left| \psi \right\rangle
_{_x }^{B_1 B_2 ,\ldots ,B_N } \left\langle \psi \right| = \rho
_x^{^{B_1 ,B_2 ,\ldots ,B_N }} $. Moreover, noting a recently
inequality presented in Ref.\cite{Yang:2007}, for a $N$-partite
state $\rho _{_x }^{B_1 B_2,\ldots ,B_N } $, we have $\frac{E_{sq}^q
\left( {\rho _x^{^{B_1 ,B_2 ,\ldots ,B_N }} } \right)}{N} \ge
K_D^{\left( N \right)} \left( {\rho _x^{^{B_1 ,B_2 ,\ldots ,B_N }} }
\right)$, where $K_D^{\left( N \right)} \left( {\rho _x^{^{B_1 ,B_2
,\ldots ,B_N }} } \right)$ denotes the distillable key of the state
$\rho _x^{^{B_1 ,B_2 ,\ldots ,B_N }} $. Thus Eq.(\ref{eq7}) can be
further written as $I^{LOCC} \le S\left( {\rho ^{B_1 }} \right) +
S\left( {\rho ^{B_2 }} \right) + \cdots + S\left( {\rho ^{B_N }}
\right) - \sum\nolimits_x {p_x K_D^{\left( N \right)} \left( {\left|
\psi \right\rangle _{_x }^{B_1 B_2 ,\ldots , B_N } } \right)} $. On
the other hand, $S\left( {\rho ^{B_1 }} \right) + S\left( {\rho
^{B_2 }} \right) + \cdots + S\left( {\rho ^{B_N }} \right) \le D$,
where $D = \log _2 d_1 d_2 \cdots d_N $, this gives the following
complementarity relation $I^{LOCC} + \sum\nolimits_x {p_x
K_D^{\left( N \right)} \left( {\left| \psi \right\rangle _{_x }^{B_1
B_2 \cdots B_N } } \right)} \le D$. This inequality shows that the
locally accessible information has close relation with the
distillable key of the state for the pure ensemble states. We
conjecture this relation also holds for the general mixed state
ensembles however we were unable to verify or disprove this
statement.

\begin{figure}[ptb]
\includegraphics[scale=0.65,angle=0]{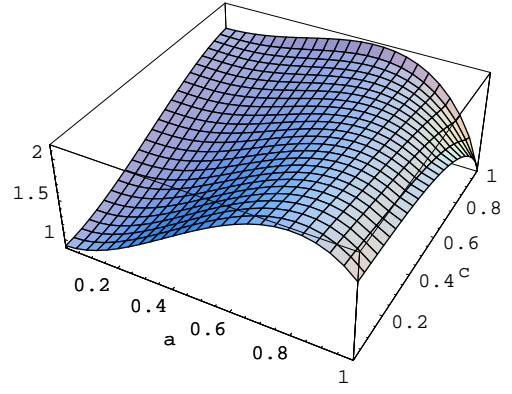}\caption{
(Color online).Plot of $I^{LOCC}$ for the ensemble \({\cal E}_1\).}
\label{fig1}%
\end{figure}

\emph{Example 1}. Consider a tripartite ensemble \({\cal E}_1\)
consisting(with equal probabilities) of the three states

\begin{equation}
\label{eq8} \left| \psi \right\rangle _{1,2} = a\left| {000}
\right\rangle \pm b\left| {111} \right\rangle , \quad \left| \psi
\right\rangle _3 = c\left| {001} \right\rangle + d\left| {110}
\right\rangle ,
\end{equation}

\noindent where we have assumed that $a,b$ and $c,d$ are both
positive real numbers with $a(c)^2 + b(d)^2 = 1$. In Fig.1, we plot
the upper bound of $I^{LOCC}$ for all values of $a$ and $c$ with
$0\le a(c)\le1$ according to Eq.(7).

\emph{Example 2}. Let us evaluate the upper bound of the locally
accessible information for the tripartite ensemble \({\cal E}_2\)
consisting(with equal probabilities) of the six states

\begin{eqnarray}
\left| \psi \right\rangle _{1,2} = a\left| {000} \right\rangle \pm
b\left| {111} \right\rangle ,  \notag \\
\left| \psi \right\rangle _{3,4} = a\left| {001} \right\rangle \pm
b\left| {110} \right\rangle ,  \notag \\
\label{eq9} \left| \psi \right\rangle _{5,6} = a\left| {010}
\right\rangle \pm b\left| {101} \right\rangle .
\end{eqnarray}

\begin{figure}[ptb]
\includegraphics[scale=0.74,angle=0]{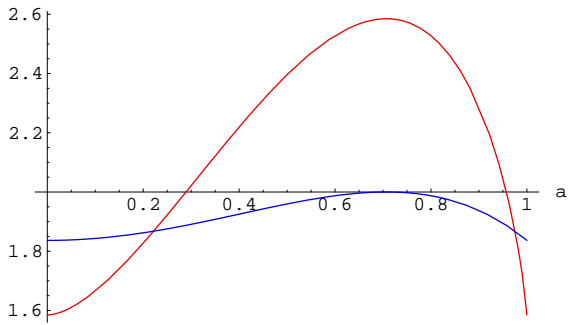}\caption{
(Color online). Plots of $I^{LOCC}$ (blue line) and $I$ (red line)
for the ensemble \({\cal E}_2\).}
\label{fig1}%
\end{figure}

Using Lemma 2, we have $I^{LOCC} \le - \frac{2}{3}\left( {1 + a^2}
\right)\log \frac{1}{3}\left( {1 + a^2} \right) - \frac{2}{3}\left(
{2 - a^2} \right)\log \frac{1}{3}\left( {2 - a^2} \right)$, on the
other hand, the ensemble \({\cal E}_2\) contains the information $I
= S\left( {\rho ^{ABC}} \right) = - a^2\log \frac{1}{3}a^2 - \left(
{1 - a^2} \right)\log \frac{1}{3}\left( {1 - a^2} \right)$. For a
vivid comparison, we plot $I^{LOCC}$ and $I$ in Fig.2. It is shown
that $I^{LOCC} < I$ whenever $0.222<a<0.975$ . Since the locally
accessible information extracted is less than the information
contained in the ensemble, it follows immediately that the
tripartite ensemble \({\cal E}_2\) consisting of the six states is
indistinguishable under LOCC if $0.222<a<0.975$.

\emph{Example 3}. Consider the following 4-partite ensemble \({\cal
E}_3\) consisting(with equal probabilities) of the nine orthogonal
states

\begin{eqnarray}
\left| \psi  \right\rangle _1  = \frac{1}{2}\left( {\left| {0000} \right\rangle  + \left| {0011} \right\rangle  + \left| {1100} \right\rangle  - \left| {1111} \right\rangle } \right),  \notag \\
 \left| \psi  \right\rangle _2  = \frac{1}{2}\left( {\left| {0000} \right\rangle  - \left| {0011} \right\rangle  + \left| {1100} \right\rangle  + \left| {1111} \right\rangle } \right),  \notag \\
 \left| \psi  \right\rangle _3  = \frac{1}{2}\left( {\left| {0001} \right\rangle  + \left| {0010} \right\rangle  + \left| {1101} \right\rangle  - \left| {1110} \right\rangle } \right),  \notag \\
 \left| \psi  \right\rangle _4  = \frac{1}{2}\left( {\left| {0001} \right\rangle  - \left| {0010} \right\rangle  + \left| {1101} \right\rangle  + \left| {1110} \right\rangle } \right),  \notag \\
 \left| \psi  \right\rangle _5  = \frac{1}{2}\left( {\left| {0101} \right\rangle  + \left| {0110} \right\rangle  + \left| {1001} \right\rangle  - \left| {1010} \right\rangle } \right),  \notag \\
 \left| \psi  \right\rangle _6  = \frac{1}{2}\left( {\left| {0101} \right\rangle  - \left| {0110} \right\rangle  + \left| {1001} \right\rangle  + \left| {1010} \right\rangle } \right),  \notag \\
 \left| \psi  \right\rangle _7  = \frac{1}{2}\left( {\left| {0111} \right\rangle  + \left| {0100} \right\rangle  + \left| {1011} \right\rangle  - \left| {1000} \right\rangle } \right),  \notag \\
 \left| \psi  \right\rangle _8  = \frac{1}{2}\left( {\left| {0111} \right\rangle  - \left| {0100} \right\rangle  + \left| {1011} \right\rangle  + \left| {1000} \right\rangle } \right),  \notag \\
 \left| \psi  \right\rangle _9  = \frac{1}{2}\left( {\left| {0000} \right\rangle  + \left| {0011} \right\rangle  - \left| {1100} \right\rangle  + \left| {1111} \right\rangle } \right).\label{eq10}
\end{eqnarray}
In this case, it is easy to show that $I^{LOCC}  \le 3$, while the
ensemble \({\cal E}_3\) contains the information $I = \log 9 >
I^{LOCC}$. Thus we conclude that ensembles \({\cal E}_3\) is
indistinguishable under LOCC.

As another application of Lemma 2, we can derive an upper bound for
the capacity of a scheme of quantum dense coding for multipartite
states. Suppose now there are $N$ Alices, say $A_1 ,A_2 ,\ldots ,A_N
$, who want to send information to $M$ receivers, Bobs, $B_1 ,B_2
,\ldots ,B_M $. They share the quantum state $\rho ^{A_1 ,A_2
,\ldots ,A_N B_1 ,B_2 ,\ldots ,B_M }$. Using the same techniques as
Ref.\cite{Bruss:2004}, we can show the capacity of distributed dense
coding is bounded by the following quantity:

\begin{equation}
\label{eq11}
\begin{array}{l}
 C\left( \rho \right) \le \log _2 d_{A_1 } + \cdots + \log _2 d_{A_N } +
S\left( {\rho ^{B_1 }} \right) + S\left( {\rho ^{B_2 }} \right) \\
 + \cdots + S\left( {\rho ^{B_M }} \right) - \mathop {\max }\limits_{Z = B_1
,B_2 ,\ldots ,B_M } \sum\nolimits_x {p_x S\left( {\rho _x^Z } \right)} . \\
 \end{array}
\end{equation}

\noindent Eq.(11) can be regarded as a generalization of the result
of Ref.\cite{Bruss:2004} to the case with multipartite senders and
multipartite receivers.

In summary, we have proposed a universal Holevo-like upper bound on
the locally accessible information for arbitrary multipartite
ensembles. This bound allows us not only to prove the
indistinguishability of some multipartite ensembles but also enables
us to obtain the upper bound for the capacity of distributed dense
coding with multipartite senders and multipartite receivers.

This work is supported by the NNSF of China, the CAS, and the
National Fundamental Research Program (under Grant No.
2006CB921900).

\end{document}